\documentclass[runningheads]{llncs}

\usepackage{graphicx}
\usepackage{amsfonts}
\usepackage{amssymb}
\usepackage{mathtools}
\usepackage{makecell}

\begin{document}

\title{Portfolio Transformer for Attention-Based\\ Asset Allocation}

\author{Damian Kisiel \and Denise Gorse}

\authorrunning{D. Kisiel and D. Gorse}

\institute{University College London, Department of Computer Science\\London, United Kingdom\\
\email{\{d.kisiel,d.gorse\}@cs.ucl.ac.uk}}

\maketitle

\begin{abstract}
Traditional approaches to financial asset allocation start with returns forecasting followed by an optimization stage that decides the optimal asset weights. Any errors made during the forecasting step reduce the accuracy of the asset weightings, and hence the profitability of the overall portfolio. The \emph{Portfolio Transformer} (PT) network, introduced here, circumvents the need to predict asset returns and instead directly optimizes the Sharpe ratio, a risk-adjusted performance metric widely used in practice. The PT is a novel end-to-end portfolio optimization framework, inspired by the numerous successes of attention mechanisms in natural language processing. With its full encoder-decoder architecture, specialized time encoding layers, and gating components, the PT has a high capacity to learn long-term dependencies among portfolio assets and hence can adapt more quickly to changing market conditions such as the COVID-19 pandemic. To demonstrate its robustness, the PT is compared against other algorithms, including the current LSTM-based state of the art, on three different datasets, with results showing that it offers the best risk-adjusted performance.

\keywords{Transformers \and Deep Learning \and Portfolio Optimization.}
\end{abstract}

\section{Introduction} \label{introduction}
Portfolio optimization algorithms aim to select the optimal weighting of financial assets in a given portfolio as a means to maximize or minimize some specific metric of interest. It is arguably the most important phase in the entire investment lifecycle, without which investors would be exposed to unacceptable levels of risk. Markowitz formally formulated this problem in what is now known as \emph{Modern Portfolio Theory} (MPT)~\cite{markowitz1952portfolio}. The risk-return trade-off pioneered by Markowitz was very influential at the time and became in effect a go-to tool for the vast majority of industry practitioners. However, despite its rigorous theoretical foundations and wide popularity, the MPT has significant shortcomings when applied in practice. One of these limitations is the assumption that future investment returns of individual assets can be predicted with a reasonable level of precision. This task has, however, been shown to be extremely difficult due to the highly stochastic nature of financial markets~\cite{michaud2008efficient}.

Recent advances in computational power and the wider availability of market data have allowed machine learning architectures to be used for portfolio optimization. For example, by using ensembles of gradient-boosted trees, one can reduce estimation errors in the returns predictions of standard Markowitz-style optimization~\cite{chen2021mean}, and the XGBoost model has been successfully used within a meta-allocation framework to switch between different risk-based strategies in order to achieve better risk-adjusted performance~\cite{10.1145/3507623.3507635}. Most recently, deep learning architectures have also started to play a major role~\cite{heaton2017deep}. However, a substantial drawback of the above portfolio selection methodologies is that they follow the classical two-step procedure in which errors in the parameter estimations of the first step are translated into inaccurate asset weightings in the second step.

Moody et al.~\cite{moody1998performance} pioneered the contrasting idea of combining prediction and performance optimization in a single step. This work was later extended by that of Zhang et al.~\cite{zhang2020deep}, who introduced an LSTM-based architecture that showed significant performance improvements over classical asset allocation techniques. In this paper, we introduce the \emph{Portfolio Transformer} (PT), which combines prediction and optimization in a novel end-to-end deep learning architecture based on an attention mechanism, directly outputting portfolio weights that optimize the Sharpe ratio, a measure of risk-adjusted return widely used in practice, under the specified transaction cost penalties. Additionally, the PT makes use of specialized gating mechanisms to determine the ideal level of non-linearity when optimizing each portfolio. Results demonstrate that the Portfolio Transformer is able to outperform a number of other methodologies, ranging from a classical optimization method to the current LSTM-based state of the art~\cite{zhang2020deep}, on three different datasets encompassing ETFs, commodities, and stocks.

\section{Background \& Related Work}

\subsection{Long Short-Term Memory (LSTM)} \label{section_LSTM}
There exists a large volume of literature applying recurrent neural networks, such as simple RNNs~\cite{tino2001financial} or Gated Recurrent Unit (GRU) networks~\cite{shen2018deep}, to financial time series prediction problems. However, the main drawback of standard recurrent neural networks, observed in multiple domains, including finance, is the so-called 'vanishing gradient problem'~\cite{hochreiter2001gradient}, whereby gradients corresponding to long-term dependencies become very small, effectively preventing the model from further training. LSTM networks~\cite{hochreiter1997long} tackle this problem by introducing gate mechanisms that allow gradients to flow unchanged; through a process of filtering and summarizing they can ignore irrelevant past information. The vanilla LSTM architecture proposed by Zhang et al.~\cite{zhang2020deep} for portfolio optimization is used in this work as one of the benchmark algorithms.

\subsection{The Transformer Model} \label{section_Transformer}
Despite their efficacy at learning time-localized patterns, LSTMs struggle to capture meaningful dependencies when the length of a sequence is relatively large. The Transformer architecture~\cite{vaswani2017attention} was developed to address this issue and has now been established as state of the art in most work in natural language processing~\cite{wolf2019huggingface}. Several studies also show the successful application of this architecture in the financial domain. For example, Wood et al.~\cite{wood2021trading} show that their Transformer model outperforms an LSTM network in time-series momentum strategies, and Xu et al.~\cite{xu2021relation} apply the model to portfolio policy learning in a reinforcement learning setting (though to maximize cumulative return rather than the industry-preferred Sharpe ratio).

At the heart of every Transformer architecture lies a mechanism called 'self-attention', which replaces recurrence and allows for simultaneous processing of all sequence elements. The Portfolio Transformer implements 'scaled dot-product attention'~\cite{vaswani2017attention}, as given by

\begin{equation} \label{eq:1}
Attention\left (Q,K,V \right) = softmax\left (\frac{QK^{T}}{\sqrt{d_{model}}} + M \right)V.
\end{equation}

\noindent The \emph{value matrix} $V\in \mathbb{R}^{\tau\times d_v}$ of equation (\ref{eq:1}) is weighted by a set of 'scores' obtained from the softmax operation, which determines how much emphasis each time step from the \emph{key matrix} $K\in \mathbb{R}^{\tau\times d_k}$ should receive when encoding sequence positions from the \emph{query matrix} $Q\in \mathbb{R}^{\tau\times d_k}$. The dot product of $Q$ and $K$ is divided by the square root of the encoding dimension ($d_{model}$) to counteract problems associated with small gradients. Additionally, the Portfolio Transformer implements masking via matrix $M\in \mathbb{R}^{\tau\times \tau}$ in the first attention block of each decoder layer to ensure it can only attend to preceding time steps and hence maintain its autoregressive property. The operation defined by equation (\ref{eq:1}) is repeated $h$ times in what is known as \emph{multi-headed attention} (MHA), 

\begin{equation} \label{eq:2}
MHA\left (Q,K,V \right) = Concatenate\left (head_1, \cdots, head_h \right)W^O,
\end{equation}

\begin{equation} \label{eq:3}
head_i = Attention\left (QW_i^Q, KW_i^K, VW_i^V \right),
\end{equation}

\noindent which allows the model to extract information from multiple representation subspaces, where each ${head}_i$ is implemented using its own set of learned linear projection matrices $W_i^Q\in \mathbb{R}^{d_{model}\times d_k}$, $W_i^K\in \mathbb{R}^{d_{model}\times d_k}$ and $W_i^V\in \mathbb{R}^{d_{model}\times d_v}$. Outputs from all attention heads are then concatenated and again linearly projected using a learned parameter matrix $W^O\in \mathbb{R}^{{hd_v}\times d_{model}}$ to obtain final values.

\subsection{Gated Residual Network (GRN)} \label{section_GRN}
In the original Transformer implementation of~\cite{vaswani2017attention} each attention layer is followed by a simple feed-forward network. The Portfolio Transformer adopts a more flexible approach and instead makes use of a \emph{Gated Residual Network} (GRN)~\cite{lim2021temporal}, acting as a gating mechanism that determines the extent of non-linear processing required for a particular portfolio, defined by

\begin{equation} \label{eq:4}
GRN\left (z \right) = LayerNorm\left (z + GLU\left (g_1 \right) \right),
\end{equation}

\begin{equation} \label{eq:5}
g_1 = W_{1}g_{2} + b_1,
\end{equation}

\begin{equation} \label{eq:6}
g_2 = ELU \left (W_{2}z + b_2 \right),
\end{equation}

\noindent in which the GRN's input is given by vector $z$, $g_1$ and $g_2$ are intermediate layers, and $ELU$ is the Exponential Linear Unit~\cite{clevert2015fast} activation function, and in which the process of filtering non-linear contributions is carried out via a Gated Linear Unit (GLU)~\cite{dauphin2017language}, which provides the Portfolio Transformer with the ability to scale down the amount of non-linear processing and default to a simpler model when, for example, the dataset is small or highly noisy.

\subsection{Time2Vec Embedding} \label{section_Time2Vec}
Since the attention layers of a Transformer do not make use of recurrence, they cannot inherently capture any information about the relative position of each element in a sequence. In the original model~\cite{vaswani2017attention} this information is injected via positional encoding. The Portfolio Transformer, however, implements the time encoding proposed by Kazemi et al.~\cite{kazemi2019time2vec}, that takes the following form:

\begin{equation} \label{eq:7}
  Time2Vec \left (t \right)[i] =
  \begin{cases}
    \omega_it + \varphi_i & \text{if $i = 0$} \\
    sin \left (\omega_it + \varphi_i \right) & \text{if 1 $\leq$ i $\leq$ k.}
  \end{cases}
\end{equation}

\noindent The temporal signal represented by $t$ in equation (\ref{eq:7}) is decomposed into a set of frequencies $\omega$ and phase shifts $\varphi$. This time decomposition technique is closely related to Fourier transforms, but instead of using a fixed set of values, all frequencies and phase shifts are learnable parameters. It should be noted that the use of $sine$ as the activation function enables the Portfolio Transformer to capture periodic behaviors in data.

\section{Methodology}

\subsection{Portfolio Transformer Architecture} \label{section_PT_architecture}
\begin{figure}
\centering
\includegraphics[scale=0.53]{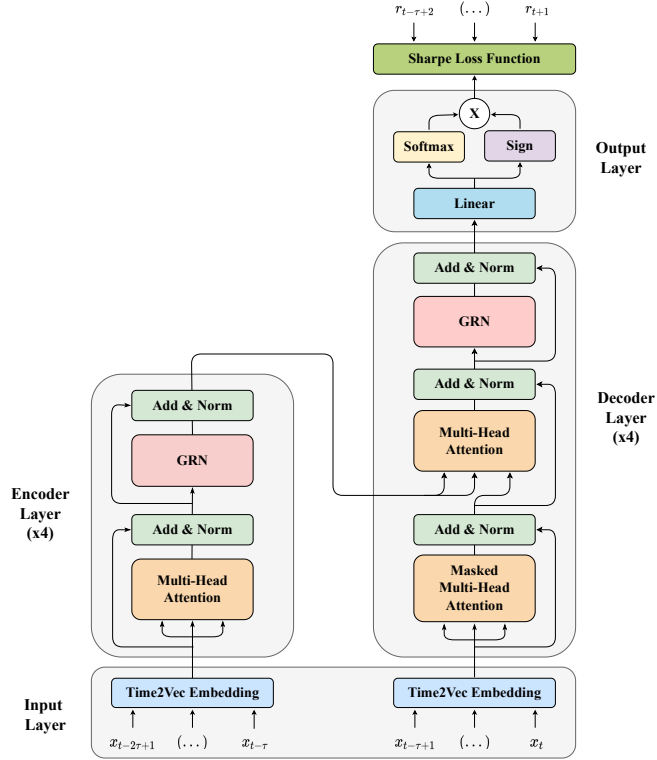}
\caption{The Portfolio Transformer: model architecture.} 
\label{fig1}
\end{figure}

The network architecture of the Portfolio Transformer is shown in Fig.~\ref{fig1}. It consists of four main building blocks: input layer (Time2Vec embedding, section~\ref{section_Time2Vec}), encoder and decoder (sections~\ref{section_Transformer} and~\ref{section_GRN}), and output layer. Each of these blocks will now be discussed in turn.

\subsubsection{Input Layer.}
Each sequence position, denoted by vector $x_t$ in Fig.~\ref{fig1}, contains concatenated returns of all $N$ assets in a given portfolio on day $t$. There are $\tau$ such vectors (per encoder and decoder block) stacked together to form the input matrix $X$ of dimension ($\tau$, $N$). Time2Vec embedding is then used to extract time-encoded cross-sectional features.

\subsubsection{Encoder.}
The Portfolio Transformer uses a stack of four identical encoder layers. Inside each of these layers, the time-encoded input is first processed by a multi-headed attention mechanism, where the number of attention heads $h$ is selected during hyperparameter optimization. A series of gating mechanisms are then applied, using a GRN module that determines the ideal amount of non-linearity. A residual connection~\cite{he2016deep}, followed by layer normalization~\cite{ba2016layer}, are additionally applied to these two sub-layers, as shown in Fig.~\ref{fig1}.

\subsubsection{Decoder.}
The decoder block is also composed of four identical layers, each of which contains two multi-headed attention modules. The first one uses masking to ensure predictions made by the Portfolio Transformer depend only on data from preceding time steps. The second one allows the decoder to attend to the output of the encoder stack, which provides a much more nuanced representation of the data and the ability to learn longer-term dependencies among portfolio assets. As in the case of the encoder, a GRN is used to remove any unnecessary complexity and there is a residual connection around each sub-layer followed by layer normalization.

\subsubsection{Output Layer.}
The Portfolio Transformer allows for short-selling via a specialized output layer that implements a compound function proposed by Zhang et al.~\cite{zhang2021universal}. First, the output of the final decoder layer is processed by a fully-connected layer. The resulting vector ($s_{i,t}$) is then used to compute final portfolio weights, given by

\begin{equation} \label{eq:8}
w_{i,t} = sign \left (s_{i,t} \right) \times softmax \left (s_{i,t} \right) \triangleq sign \left (s_{i,t} \right) \times \frac{e^{s_{i,t}}}{\sum_{j=1}^{N}e^{s_{j,t}}}.
\end{equation}

\noindent The use of the softmax operation in equation (\ref{eq:8}) ensures that, while portfolio weights can be either positive or negative (the latter allowing for short-selling), the sum of their absolute values always remains equal to one.

\subsection{Loss Function} \label{section_loss_function}
The objective of the PT model as currently implemented (other objective functions being possible) is to learn the asset distribution that maximizes risk-adjusted returns as measured by the Sharpe ratio, which is defined below as expected portfolio return divided by its volatility:

\begin{equation} \label{eq:9}
SR = \frac{E(R_P)}{\sqrt{E(R_{P}^{2}) - (E(R_P))^2}}.
\end{equation}

\noindent Since transaction costs can significantly diminish the performance of allocation strategies with high turnover, the Portfolio Transformer uses cost-adjusted portfolio returns 

\begin{equation} \label{eq:10}
R_{P,t} = \sum_{i}^{N}w_{i,t-1} \times r_{i,t} - C \times \sum_{i}^{N} \left | w_{i,t-1} - w_{i,t-2} \right |
\end{equation}

\noindent in order to find solutions that account for trading costs, where $C$ is a constant cost rate, set to a realistic value of two basis points (2 bps), $w_{i,t-1}$ represents the weight of asset $i$ on day $t-1$, and $r_{i,t}$ denotes the realized arithmetic return of asset $i$ from day $t-1$ to day $t$, computed using asset prices $P_{i,t}$ and $P_{i,t-1}$ as follows:

\begin{equation} \label{eq:11}
r_{i,t} = \frac{P_{i,t}}{P_{i,t-1}} - 1.
\end{equation}

\noindent The expected portfolio return, denoted by $E\left(R_P\right)$ in equation (\ref{eq:9}), is obtained by taking an average of all portfolio returns over a trading period of length $\tau$:

\begin{equation} \label{eq:12}
E \left (R_P \right) = \frac{1}{\tau} \sum_{t=1}^{\tau} R_{P,t}.
\end{equation}

\noindent Finally, since the PT allows for short-selling but no leverage, the portfolio positions are constrained by $w_{i,t}\in[-1, 1]$ and $\sum_{i}^{N}\left|w_{i,t}\right|=1$, which is achieved through the use of the compound function in equation (\ref{eq:8}).

\subsection{Training \& Model Calibration} \label{section_training}
Data is split into training and test segments using an expanding window approach, where initially all data points before the end of 2015 are used for training and the out-of-sample test is carried out on observations recorded in 2016. The training window is then extended to include the year 2016 and the model is tested on the subsequent year (2017), and so on. This way, the model is retrained every year, with all available historical data being used to update the network parameters. Portfolio positions are adjusted on a daily basis, and a transaction cost rate of 2 bps is used during performance evaluation.

The PT network is trained via mini-batch stochastic gradient descent (with batch size being among the hyperparameters) using the Adam optimizer~\cite{kingma2014adam}. For model calibration purposes, and to control for overfitting, 10\% of any training segment is set aside as a separate validation set. Hyperparameter optimization, conducted using 100 iterations of random grid search, is performed only on the validation set, ensuring that the model has access to test data only during the performance evaluation stage. To further improve the model's ability to generalize, early stopping is implemented. The Portfolio Transformer was developed using the TensorFlow framework and all experiments were conducted on NVIDIA's Tesla P100 16GB GPU with 55GB of RAM memory.

\subsection{Datasets Used} \label{section_datasets}
The efficacy of the Portfolio Transformer is demonstrated on three datasets containing daily price observations. The first of these datasets starts in 2006 and is composed of the same four Exchange Traded Funds (ETFs) used in the LSTM-based experiments carried out by Zhang et al.~\cite{zhang2020deep}: AGG (aggregate bond index), DBC (commodity index), VIX (volatility index), and VTI (US stocks index). The second dataset, which starts in 2002, is composed of 24 continuous commodity futures contracts, including metals, agricultural products, and energy commodities such as oil and natural gas. Finally, the PT model is tested on daily observations of 500 stocks based in the US and listed on NASDAQ. This last dataset starts in 1996 and aims to demonstrate the model's performance on a large portfolio of hundreds of instruments.

\subsection{Benchmark Models} \label{section_benchmarks}
Four algorithms are implemented as benchmarks: (1) mean-variance optimization (MV)~\cite{markowitz1952portfolio}, a classical two-step portfolio selection procedure with a moving window of 50 days used to estimate expected asset returns and covariances; (2) XGBoost~\cite{chen2016xgboost}, included because gradient-boosted decision trees perform very well in applied machine learning competitions; (3) a multilayer perceptron (MLP), as a universal function approximator that can capture highly non-linear dependencies; and (4) the LSTM architecture of~\cite{zhang2020deep}, a high-performance recurrent architecture that represents the previous state of the art.

\section{Results}

\subsection{Performance Comparison: Full Investment Horizon} \label{section_performance_comparison}
The performance of the Portfolio Transformer is compared to that of the benchmark algorithms of section \ref{section_benchmarks} using a number of metrics that aim to capture portfolio risk level through annualized volatility (Vol.) and maximum drawdown (MDD), profitability via annualized returns (Returns) and percentage of positive returns (\% of + Ret), and risk-adjusted performance using annualized Sharpe, Sortino and Calmar ratios.

\begin{table}
\centering
\caption{Experimental results for different algorithms and datasets.}
\label{tab1}
\begin{tabular}{lccccccc}
\hline
 &  \bfseries Returns & \bfseries Vol. & \bfseries Sharpe & \bfseries Sortino & \bfseries MDD & \bfseries Calmar & \makecell{\bfseries \% of \\ \bfseries +Ret} \\
\hline
\bfseries Panel A: ETFs &  &  &  &  &  &  &  \\
\hline
MV & 0.004 & 0.122 & 0.012 & 0.270 & 0.120 & 0.836 & 0.497 \\
XGBoost & 0.100 & 0.140 & 0.657 & 1.240 & 0.116 & 2.591 & 0.496 \\
MLP & 0.135 & 0.128 & 0.923 & 1.789 & 0.087 & 2.225 & 0.498 \\
LSTM & \bfseries 0.215 & 0.133 & 1.539 & 2.830 & 0.096 & 3.621 & 0.535 \\
Portfolio Transformer & 0.138 & \bfseries 0.067 & \bfseries 2.252 & \bfseries 4.093 & \bfseries 0.036 & \bfseries 4.773 & \bfseries 0.548 \\
\hline
\bfseries Panel B: Commodities &  &  &  &  &  &  &  \\
\hline
MV & 0.008 & 0.053 & 0.174 & 0.342 & 0.059 & 0.682 & 0.519 \\
XGBoost & 0.015 & 0.059 & 0.277 & 0.420 & 0.063 & 0.267 & 0.505 \\
MLP & 0.026 & 0.056 & 0.479 & 0.727 & 0.057 & 0.797 & 0.515 \\
LSTM & 0.038 & \bfseries 0.031 & 1.182 & 1.852 & \bfseries 0.023 & 2.108 & 0.528 \\
Portfolio Transformer & \bfseries 0.174 & 0.108 & \bfseries 1.506 & \bfseries 2.304 & 0.077 & \bfseries 2.272 & \bfseries 0.543 \\
\hline
\bfseries Panel C: Stocks &  &  &  &  &  &  &  \\
\hline
MV & 0.079 & 0.126 & 0.694 & 1.126 & 0.106 & 1.386 & 0.523 \\
XGBoost & 0.101 & 0.118 & 0.923 & 1.352 & 0.080 & 1.491 & 0.533 \\
MLP & 0.089 & 0.121 & 0.767 & 1.102 & 0.087 & 1.236 & 0.534 \\
LSTM & 0.111 & \bfseries 0.077 & 1.456 & 2.155 & \bfseries 0.056 & 2.561 & 0.565 \\
Portfolio Transformer & \bfseries 0.334 & 0.147 & \bfseries 2.001 & \bfseries 3.440 & 0.091 & \bfseries 4.824 & \bfseries 0.566 \\
\hline
\end{tabular}
\end{table}

\begin{figure}
\centering
\includegraphics[scale=0.07]{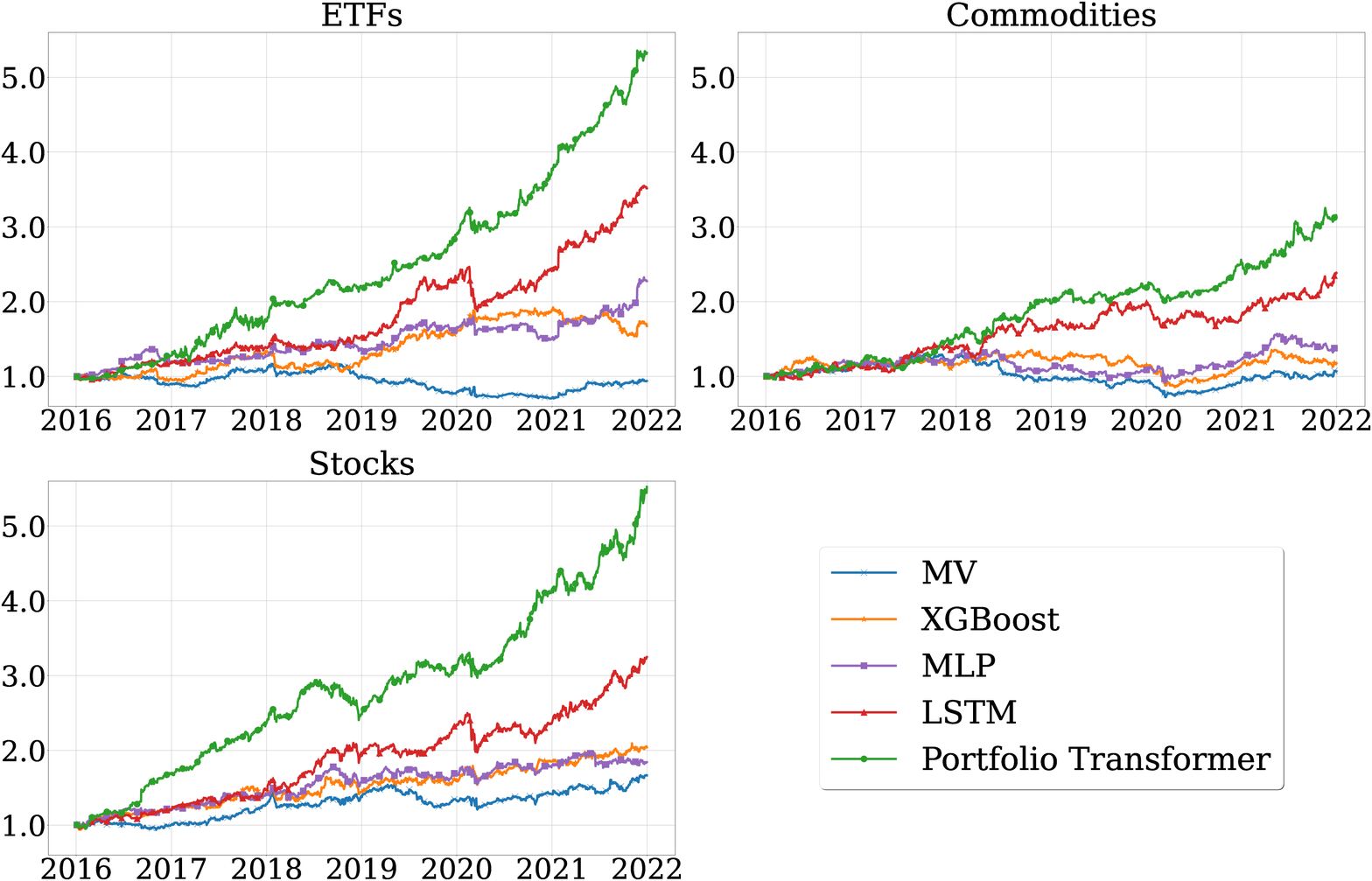}
\caption{Comparison of cumulative returns.} 
\label{fig2}
\end{figure}

\noindent Panel A in Table~\ref{tab1} shows the results for the portfolio of ETFs, where the Portfolio Transformer outperforms the benchmarks on all but one metric, annualized returns. These are higher for the LSTM model, but this comes at a cost of increased volatility; when this is accounted for, by using the Sharpe ratio, the PT offers much higher \emph{risk-adjusted} returns. Results presented in Panels B (commodities) and C (stocks) show that in both cases the PT achieves the best risk-adjusted performance, while also delivering the highest annualized returns. The PT is beaten by the LSTM in these cases on only two out of seven metrics, volatility and maximum drawdown. In the former case, this is compensated by the PT's higher Sharpe ratio and in the latter by the higher Calmar ratio, relative to the LSTM; the Sharpe ratio is a measure of portfolio return adjusted by volatility, and Calmar ratio of portfolio return adjusted by maximum drawdown, arguably of more relevance than volatility and maximum drawdown per se.

\noindent The cumulative return plots of Fig.~\ref{fig2} demonstrate the superior performance of the Portfolio Transformer over the whole investment horizon. The PT generates the highest cumulative returns and offers a reasonable risk profile for all three datasets. The second-best performing model is the LSTM, suggesting that time dependencies learned through recurrence, in case of the LSTM model, or through an attention mechanism, in case of the PT, are very useful in a portfolio optimization setting. However, it can be seen that the LSTM model struggles during the COVID-19 crisis (first quarter of 2020), while the attention-driven Portfolio Transformer shows a much quicker response to this sudden market regime change. The performances of the XGBoost and MLP models are comparable, but lag considerably behind those of the LSTM and the PT, while mean-variance optimization (MV) is by far the worst-performing algorithm, suggesting that highly inaccurate asset weightings are generated by this classical two-step procedure. 

\subsection{Performance Comparison: COVID-19 Crisis}

\begin{figure}
\centering
\includegraphics[scale=0.19]{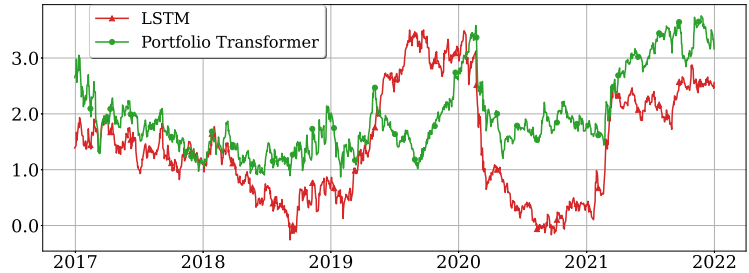}
\caption{Rolling Sharpe ratio (12-month) on the ETF dataset.} 
\label{fig3}
\end{figure}

The above-mentioned difference between the performances of the PT and LSTM models during this period of extreme market volatility is further illustrated in Fig.~\ref{fig3}, which shows the 12-month rolling Sharpe ratio of the two models on the ETF dataset. The LSTM suffers a large fall in its risk-adjusted returns during this period and there are times when it even drops down below zero. The Portfolio Transformer, on the other hand, shows a more stable behavior, with its rolling Sharpe ratio staying mostly well above one and delivering outstanding risk-adjusted performance during the Bull market that followed the crisis.

\section{Conclusions}
This work has introduced the Portfolio Transformer (PT), which directly optimizes risk-adjusted returns using a novel end-to-end attention-based architecture with specialized time-encoding layers and gating mechanisms. By incorporating transaction costs directly into its loss function the PT model can account for trading cost constraints faced by investors. The results demonstrate that the PT model delivers exceptional risk-adjusted performance, in this respect outperforming all benchmark algorithms on three different datasets with varying portfolio sizes. Due  to its full encoder-decoder configuration and its attention mechanism the Portfolio Transformer is able to learn long-term dependencies and as a result can react more quickly to changing market regimes, as demonstrated by its response to the COVID-19 crisis. Turning to future work, one extension of the current model could study the Portfolio Transformer's performance under an attention mechanism different from the scaled dot-product attention currently used. In addition, while it is popular with industry practitioners, the Sharpe ratio is only one of many possible objective functions that could be optimized, and subsequent work will consider alternative metrics and their effect on the overall portfolio performance.


\bibliographystyle{splncs04}
\bibliography{refs}

\end{document}